\documentclass[english,showpacs,floatfix,preprint,12pt]{revtex4}

\usepackage[dvips]{graphicx}
\usepackage{longtable}
\usepackage{amsmath,amssymb}
\usepackage{dcolumn}
\usepackage{latexsym}
\usepackage{babel}
\usepackage[latin1]{inputenc}
\usepackage{color}
\usepackage[colorlinks]{hyperref}

\begin{document}
\title{Observational constrains on the DGP brane-world model with a Gauss-Bonnet
 term in the bulk}

\author{Jian-Hua He, Bin Wang}
\email{wangb@fudan.edu.cn} \affiliation{Department of Physics,
Fudan University, 200433 Shanghai, China}

\author{Eleftherios Papantonopoulos}
\email{lpapa@central.ntua.gr} \affiliation{Department of Physics,
National Technical University of Athens, GR 157 73 Athens, Greece}

\begin{abstract}
Using the data coming from the new 182 Gold type Ia supernova
samples, the baryon acoustic oscillation measurement from the
Sloan Digital Sky Survey and the H(z) data, we have performed a
statistical joint analysis of the DGP brane-world model with a
high curvature Gauss-Bonnet term in the bulk. Consistent
parameters estimations show that the Gauss-Bonnet-Induced Gravity
model is a viable candidate to explain the observed acceleration
of our universe.
\end{abstract}
\maketitle

A variety of cosmological observations suggests a concordant
compelling result that our universe is undergoing an accelerated
expansion, which is one of the deepest theoretical problems in
cosmology~\cite{1}. Within the framework of general relativity,
the acceleration must be associated with the so called dark
energy, whose theoretical nature and origin are the source of much
debate. Despite the effective negative equation of state
$\omega<-1/3$ from the robust observational evidence, we know
little about the dark energy.

An alternative approach which does not need dark energy to explain
the late-time acceleration is motivated by sting theory via the
brane-world scenarios. In the late-time universe,  one of the
simplest extra-dimensional brane-world model which describes the
cosmological evolution at low energies is the DGP model
~\cite{2},~\cite{3}. In this model, gravity leaks  off the
4-dimensional brane into the 5-dimensional bulk at large scales.
Gravity leakage at late-times initiates acceleration due to the
weakening of gravity on the brane, without the need of introducing
the mystery dark energy.

 However, the DGP model which modifies Einstein's General
Relativity in a consistent manner in the infra-red is not free of
problems. The most serious one is that such modified theories suffer
from classical and/or quantum instabilities, at least at the level
of linear perturbations. Most candidate braneworld models,  have
been shown to suffer from such instabilities or strong coupling or
both, \cite{DGPghosts,strong}. Generically, a ghost mode appears in
the perturbative spectrum of the theory at the scale where gravity
is modified, effectively driving the acceleration. Therefore some
kind of ultra-violet completion is needed for the DGP model in order
to be safe at strong coupling.

There have been some attempts to generalize the DGP model so that
they can show ultra-violet modifications to General Relativity.
One possible way is to introduce a high curvature Gauss-Bonnet
(GB) term in the gravitational action to display the higher energy
stringy corrections~\cite{5},~\cite{6}. An intriguing cosmological
model with the combination of infra-red and ultra-violet
modifications by introducing the GB term in the 5D Minkowski bulk
containing a Friedmann brane with DGP induced gravity, was
presented in~\cite{7}. In the general GB correction to the Induced
Gravity, the late-time self acceleration of the universe is still
kept, and striking new behaviour in the early universe is also
shown~\cite{Kofinas:2003rz,bb}. It is of great interest to
investigate whether such model is a viable cosmological model.

The pure DGP model was tested using  data from various
observations ~\cite{DGPobservations,DGPobservations1,8,11,12,00}.
In this work we are going to impose constraints on the model
parameters by using the latest SNIa data compiled by Riess et
al~\cite{9}, the baryon acoustic oscillations (BAO) measurement
from the large-scale correlation function of the Sloan Digital Sky
Survey(SDSS) luminous red galaxies \cite{10} in combination with
the H(z) data. We will compare our results with the cosmological
consequences of the DGP model as they were discussed
in~\cite{8},~\cite{11},~\cite{12} to see the influence of the GB
effect on the DGP model and also disclose the value of the GB
parameter from  observations. The GB correction term has been
found effective on the modification of the cosmological evolution
around $z\sim 1$. This has been reported, for example, on the
influence of the equation of state either in the modified RS model
or the modified DGP model \cite{b1,b2}.

All the tests we will use to constrain the parameters of our model
are for relatively low redshift data. It would be interesting to
test our model for high redshifts using observational data from
CMB anisotropies and matter power spectrum. However, this would
require the knowledge of evolution of density perturbations of our
model, a subject which is not fully understood even in the pure
DGP model \cite{DGPobservations,a1}.

Combing the GB term in the bulk with the Induced Gravity on the
brane, the Friedmann equation on the DGP brane can be described by
the dimensionless variables in the form of~\cite{7}
\begin{equation}
  4(\gamma h^2+1)h^2=(h^2-\mu)^2~, \label{1}
\end{equation}
where the dimensionless variables are
$\gamma=\frac{8\alpha}{3r^2},~
h=Hr,~\mu=\frac{r\kappa_5^2}{3}\rho$. The conservation equation
becomes
\begin{equation}
  \mu'+3h(1+\omega)\mu=0~,\label{2}
\end{equation}
where the prime denotes $d/d\tau$ and $\omega$ is the equation of
state. Here $r$ is the crossover length scale, which is two times
of the value defined in ~\cite{8},~\cite{11},~\cite{12} where
$r_c=M_4^2/2M_5^2$ and $\alpha$ is the GB coupling constant which
has the dimensions of $(lenght)^2$.

As discussed in ~\cite{7}, the physically relevant
self-accelerating solution which is the generalization of the DGP
model exists when $0<\gamma<1/16$ and has the Friedmann equation
\begin{equation}
  H^2=\frac{1-8\gamma}{12\gamma^2
  r^2}+\frac{\sqrt{(1-8\gamma)^2-8\gamma^2(3\mu_z+6)}}{6\gamma^2r^2}cos\left(\theta_z+\frac{4\pi}{3}\right)~,\label{3}
\end{equation}
\begin{equation}
 cos3\theta_z=\frac{216\gamma^4\mu_z^2-36\gamma^2(1-8\gamma)(\mu_z+2)+(1-8\gamma)^3}{[(1-8\gamma)^2
  -24\gamma^2(\mu_z+2)]^{3/2}}~. \label{4}
\end{equation}
In the special case where $\alpha=0$~($\gamma=0$),  the (+) branch
of the DGP model can be recovered with the Friedmann equation
\begin{equation}
\lim_{\gamma \to
0^{+}}H^2=\frac{2}{r}H+\frac{\kappa_5^2}{3r}\rho_z~. \label{6}
\end{equation}
For the benefit of the following discussion, we rewrite
Eq~(\ref{3})  in the form
\begin{equation}
  E^2(z)=\frac{H^2}{H_0^2}=\frac{1-8\gamma}{12\gamma^2
  h_0^2}+\frac{\sqrt{(1-8\gamma)^2-8\gamma^2(3\mu_z+6)}}{6\gamma^2h_0^2}cos\left(\theta_z+\frac{4\pi}{3}\right)~,
  \label{6}
\end{equation}
where $\mu_z=\mu_0(1+z)^3$. When $z=0$ we arrive at
\begin{equation}
  h_0^2=\frac{1-8\gamma}{12\gamma^2}+\frac{\sqrt{(1-8\gamma)^2-8\gamma^2(3\mu_0+6)}}{6\gamma^2}cos\left(\theta_0+\frac{4\pi}{3}\right)~,\label{7}
\end{equation}
\begin{equation}
 cos3\theta_0=\frac{216\gamma^4\mu_0^2-36\gamma^2(1-8\gamma)(\mu_0+2)+(1-8\gamma)^3}{[(1-8\gamma)^2
  -24\gamma^2(\mu_0+2)]^{3/2}}~,\label{8}
\end{equation}
where $h_0=H_0r$. Neglecting the GB correction ($\gamma\rightarrow
0$) Eq~(\ref{6}) reduces to
\begin{equation}
\lim_{\gamma \to
0^{+}}E(z)^2=\left[\sqrt{\Omega_{m0}(1+z)^3+4\Omega_r}+\sqrt{4\Omega_r}\right]^2~,
\label{9}
\end{equation}
where
$\Omega_{m0}=\frac{\kappa_4^2\rho}{3H_0^2}=\frac{\mu_0}{h_0^2},~
\mu_0=\frac{\Omega_{m0}}{4\Omega_r},~ \Omega_r=\frac{1}{4h_0^2}$.
 In our notation which is consistent with that used in
~\cite{7}, the crossover factor $r=2r_c$, which leads to
$\Omega_r=1/4\Omega_{rc}$. Then Eq~(\ref{9}) can go back to the
equation in the pure DGP model described in ~\cite{8,11,12}. Due
to the GB correction, the cross-over scale obeys~\cite{7}
\begin{equation}
2H_0^{-1}\leq r\leq 4H_0^{-1}~,\label{10}
\end{equation}
while in the DGP(+) limit ($\gamma \rightarrow 0 $), $r \sim
2H_0^{-1}$. It was found that the physically relevant
self-accelerating solution of the GB correction to the Induced
Gravity has a finite temperature big bang, since the density
$\mu_z$ is bounded from above~\cite{Kofinas:2003rz}. This upper
bound is also the requirement of real value of the square root in
Eq~(\ref{4}) and Eq~(\ref{6}) which is
\begin{equation}
\mu_z\leq \mu_{\rm
max}'=\frac{1-16\gamma+16\gamma^2}{24\gamma^2}~.\label{11}
\end{equation}
Requiring $|cos3\theta_z|<1$, (there is a milder condition:
$\gamma<1/16$ ) we can have
\begin{equation} h_{\rm
max} =\frac{1+\sqrt{1-12\gamma}}{6\gamma}~,\label{12}
\end{equation}
which is the initial Hubble rate for the model. Then the upper
bound density of GB corrected Induced Gravity model reads
\begin{equation}
\mu_z<\mu_{\rm max} = \frac{1}{3}h_{max}^2(2\sqrt{1-12\gamma}-1)<
\mu_{\rm max}'~,\label{13}
\end{equation}
where $\mu_{max}$ is the initial density with $\gamma<1/16$. If
$\gamma\rightarrow 0$, DGP model will be restored and
$h_{max}=\mu_{max}=\infty$. { Using the Induced Gravity model with
the GB term of the bulk to describe the physically relevant
self-acceleration, the density upper bound indicates that our
universe started from a finite redshift $z_{max}$ instead of a
singularity at $z=\infty$. For the universe without dark energy,
$\mu_z=\mu_0(1+z)^3$ the upper bound on the density leads to
\begin{equation}
(1+z_{\rm obs})^3<(1+z_{\rm max})^3=\frac{\mu_{\rm
max}}{\mu_0}~,\label{14}
\end{equation}
where $z_{obs}$ is the redshift we have the observational data and
$z_{max}$ is the starting moment of the universe in this model.

In the following we are going to constrain  this model by using
the latest observational data, such as the gold SN Ia data, the
BAO measurement from SDSS and combing these obsevations with H(z)
data.

The up-to-date gold SN Ia sample was compiled by Riess et
al~\cite{9}. This sample consists of 182 data, in which 16 points
with $0.46<z<1.39$ were obtained recently by the Hubble Space
Telescope(HST), 47 points with $0.25<z<0.56$ by the first year
Supernova Legacy Survey(SNLS) and the remaining 119 points are old
data. The SN Ia observation gives the distance modulus of SN at
the redshift $z$. The distance modulus is defined as
\begin{equation}
\mu\equiv m-M=\log_{10}D_L+5\log_{10} \left(\frac{1/H_0}{\rm
Mpc}\right)+25~,\label{15}
\end{equation}
where  $D_{L}$ is the dimensionless luminosity distance and it given
by }$D_{L}=(1+z)\int_0^z\frac{dz'}{E(z')}$~.

\begin{figure}
\includegraphics[width=3.5in,height=2.9in]{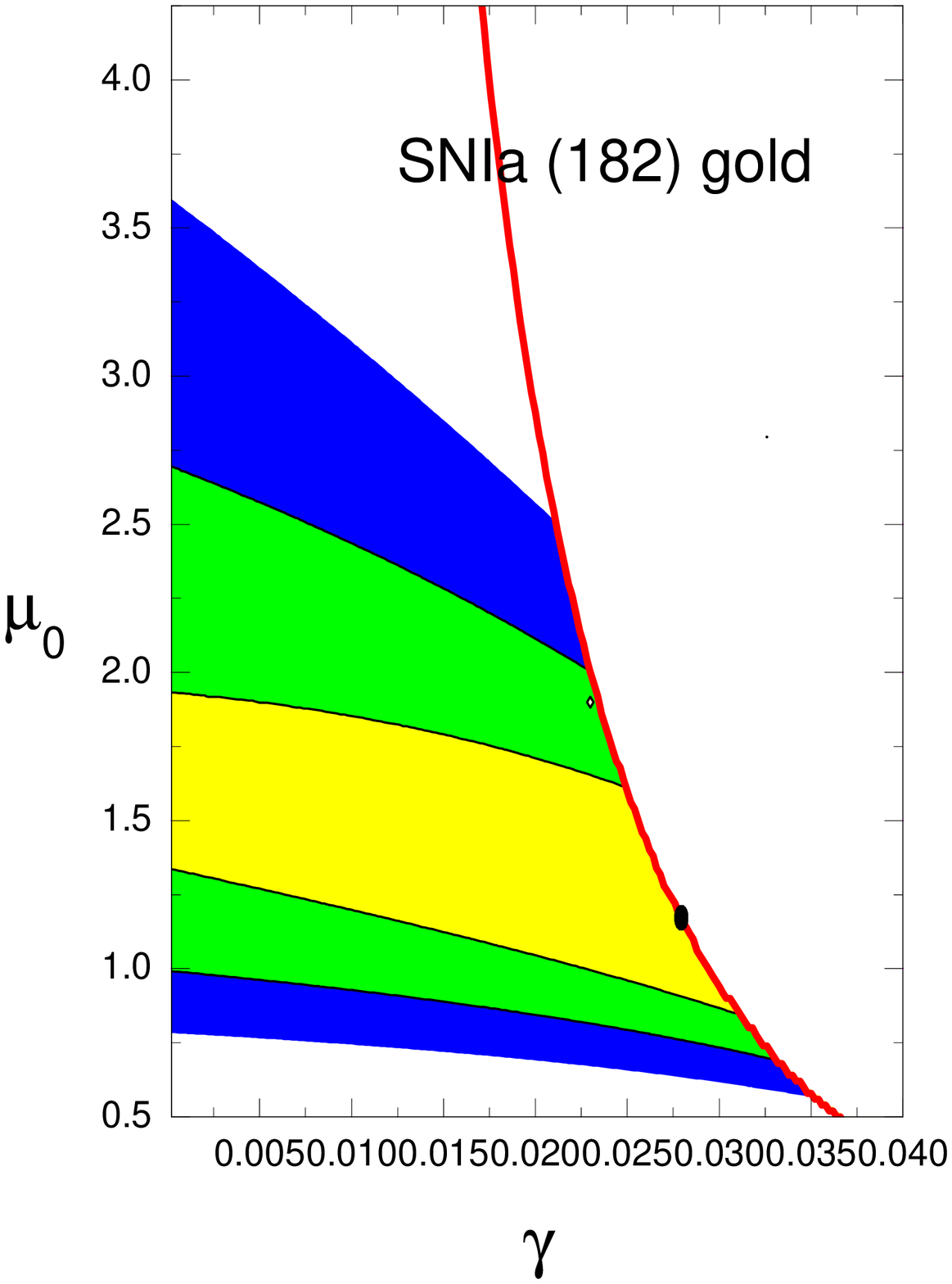}
\includegraphics[width=2.8in,height=2.8in]{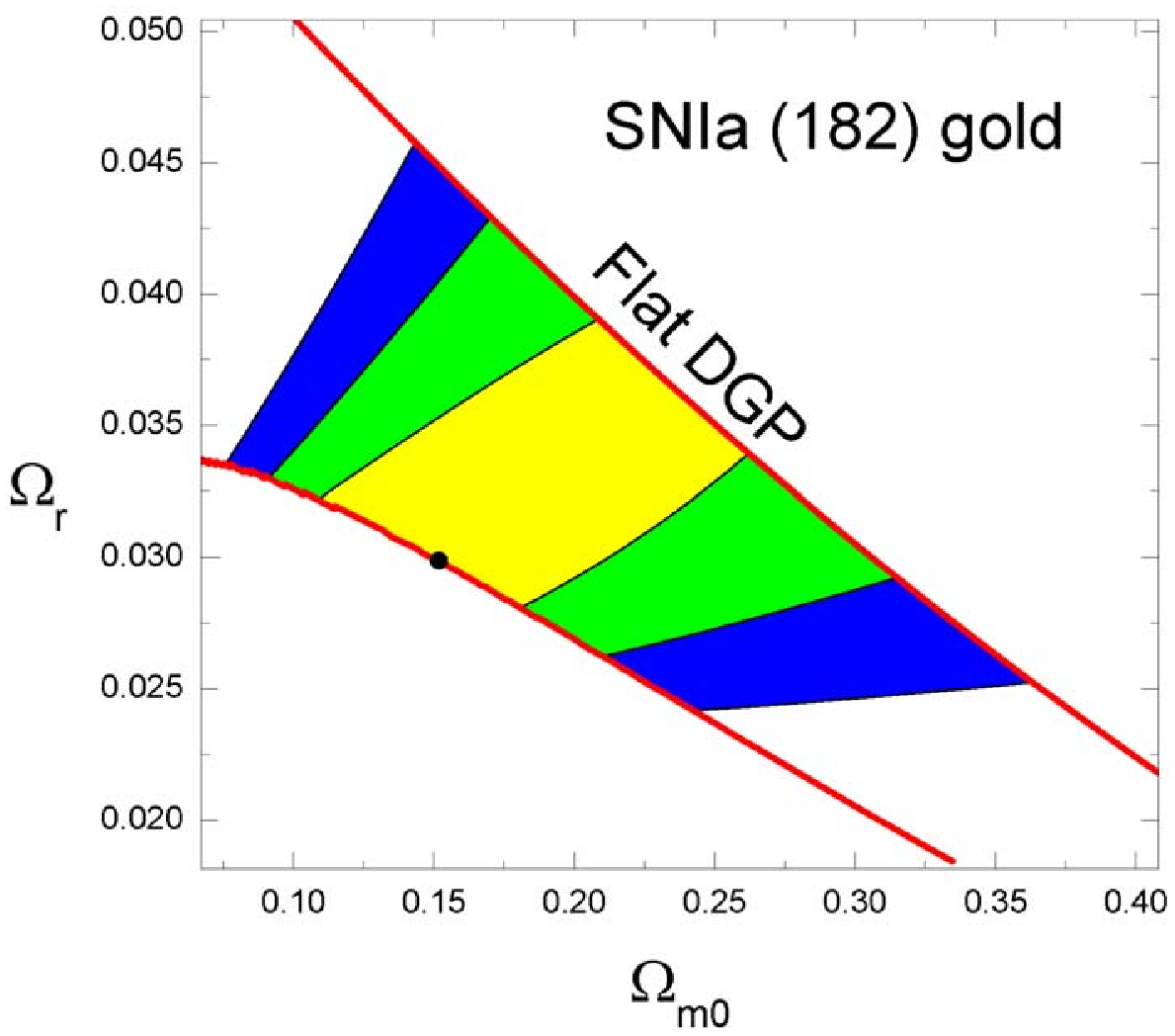}
\caption{ Probability contours at 68.3\%, 95.4\% and 99.7\%
confidence levels for joint parameters: (a) ($\gamma $,
$\mu_{m0}$)-plane from the gold sample of SNIa data. The point
shows the best-fit from the SNIa analysis. The red line
corresponds to the limit due to Eq~(\ref{13}). (b) For
($\Omega_{m0}$, $\Omega_{r}$)-plane.
The upper boundary corresponds to $\gamma=0$ (pure DGP) and the lower boundary is due to Eq~(\ref{13}).  \\
\label{one}}
\end{figure}

From Eq~(\ref{6}) we see that there are two parameters
$\mu_0$,~$\gamma$ in the model. Eq~(\ref{7}) tells us that $h_0$
is a function of $\mu_0$ and $\gamma$. In order to place
constraints on the model, we perform $\chi^2$ statistics for the
model parameter
\begin{equation}
\chi^2_{SN}(\gamma,\mu_0,\overline{M})=\sum_{i}\frac{|\mu_{obs}(z_i)-\mu_{th}(z_i)|^2}{\sigma_i^2}~.
\end{equation}
The best-fit values of parameter are shown in Table~\ref{t1},
where we have done the  marginalization of the nuisance parameter
$\overline{M} = 5\log_{10} \left(\frac{1/H_0}{\rm Mpc}\right)+25$.
 With the best-fit values of $\mu_0$,~$\gamma$ we can get other
 parameters $h_0$,~$\alpha$ from their relations, which
are also listed in Table~\ref{t1}. In Fig~\ref{one}(a), we present
the contours of 68.3\%, 95.4\% and 99.7\% confidence levels. It is
of interest to disclose parameters such as $\Omega_{m0}$,
$\Omega_{r}$ which have direct physical meanings so that we can
compare our model with the pure DGP model. Recalling the relation
between ($\mu_0$, $\gamma$) and ($\Omega_{m0}$, $\Omega_{r}$) and
noting that the parameters transformations from ($\mu_0$,
$\gamma$) to ($\Omega_{m0}$, $\Omega_{r}$) have non-zero Jacobi
determinant $\frac{\partial(\Omega
_{m0},\Omega_{r})}{\partial(\gamma,\mu_0)}$, we can obtain the
constraint on the physical parameters ($\Omega_{m0}$,
$\Omega_{r}$) from the SNIa observations. The best-fit values are
listed in Table~\ref{t1} and contours are shown in
Fig~\ref{one}(b). Our analysis shows that if we use  only the SNIa
data, the constrains are not good and the $1\sigma$ range is
large.
\begin{table*}
\begin{center}
\caption{} \label{t1}
\begin{tabular}{|c|c|c|c|c|c|c|c|}
  \hline
  \hline
    Test& $\gamma$ & $\mu_0$ & $\Omega_{m0}$ & $\Omega_{r}$ & $h_0$ & $\alpha$ (in $H_0^{-2}$unit) & $\chi_{min}^2$ \\
  \hline
   SNIa &$0.0278_{-0.0278}^{+0.0033}$ &$1.20_{-0.34}^{+0.70}$ &$0.15_{-0.04}^{+0.11}$  &$0.0302_{-0.0020}^{+0.0087}$ &$2.88_{-0.34}^{+0.10}$  &$0.086_{-0.086}^{+0.005}$  &158.27 \\
   SNIa+LSS& $0.000_{-0.000}^{+0.005}$ & $2.26_{-0.42}^{+0.55}$  & $0.29_{-0.03}^{+0.04}$ & $0.0318_{-0.0031}^{+0.0029}$  & $2.81_{-0.12}^{+0.15}$  & $0.000_{-0.000}^{+0.017}$& 162.92\\
   SNIa+LSS+H(z)& $0.000_{-0.000}^{+0.003}$  & $2.03_{-0.33}^{+0.40}$  & $0.27_{-0.03}^{+0.03}$ & $0.0333_{-0.0025}^{+0.0024}$ &$ 2.74_{-0.10}^{+0.11}$ &$0.000_{-0.000}^{+0.009}$ &174.04 \\
  \hline
  \hline
\end{tabular}
\end{center}
%\tablenotetext{$\alpha$ is in $H_0^{-2}$unit}
\end{table*}

\begin{figure}
\includegraphics[width=3.5in,height=3.0in]{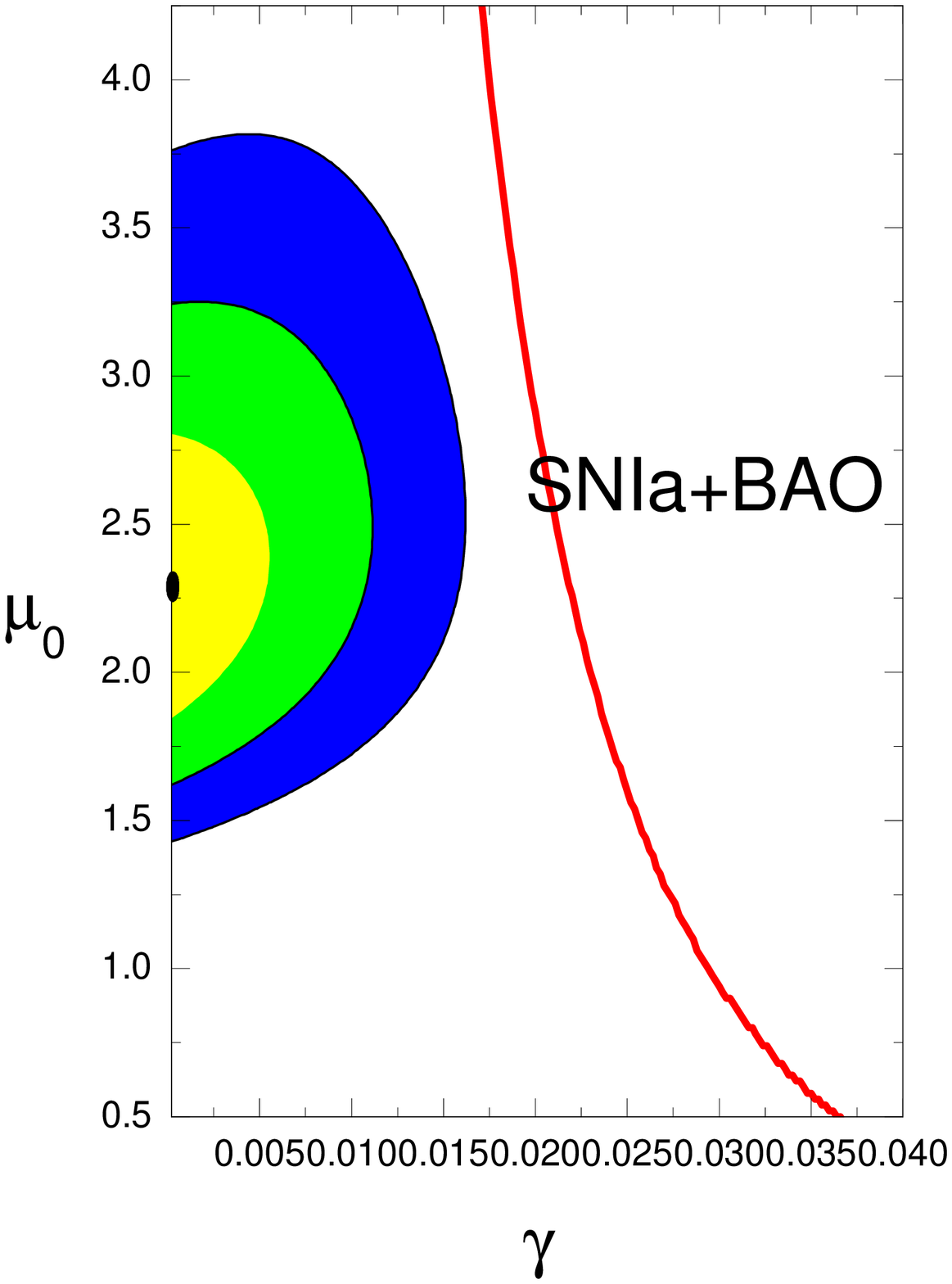}
\includegraphics[width=2.8in,height=2.8in]{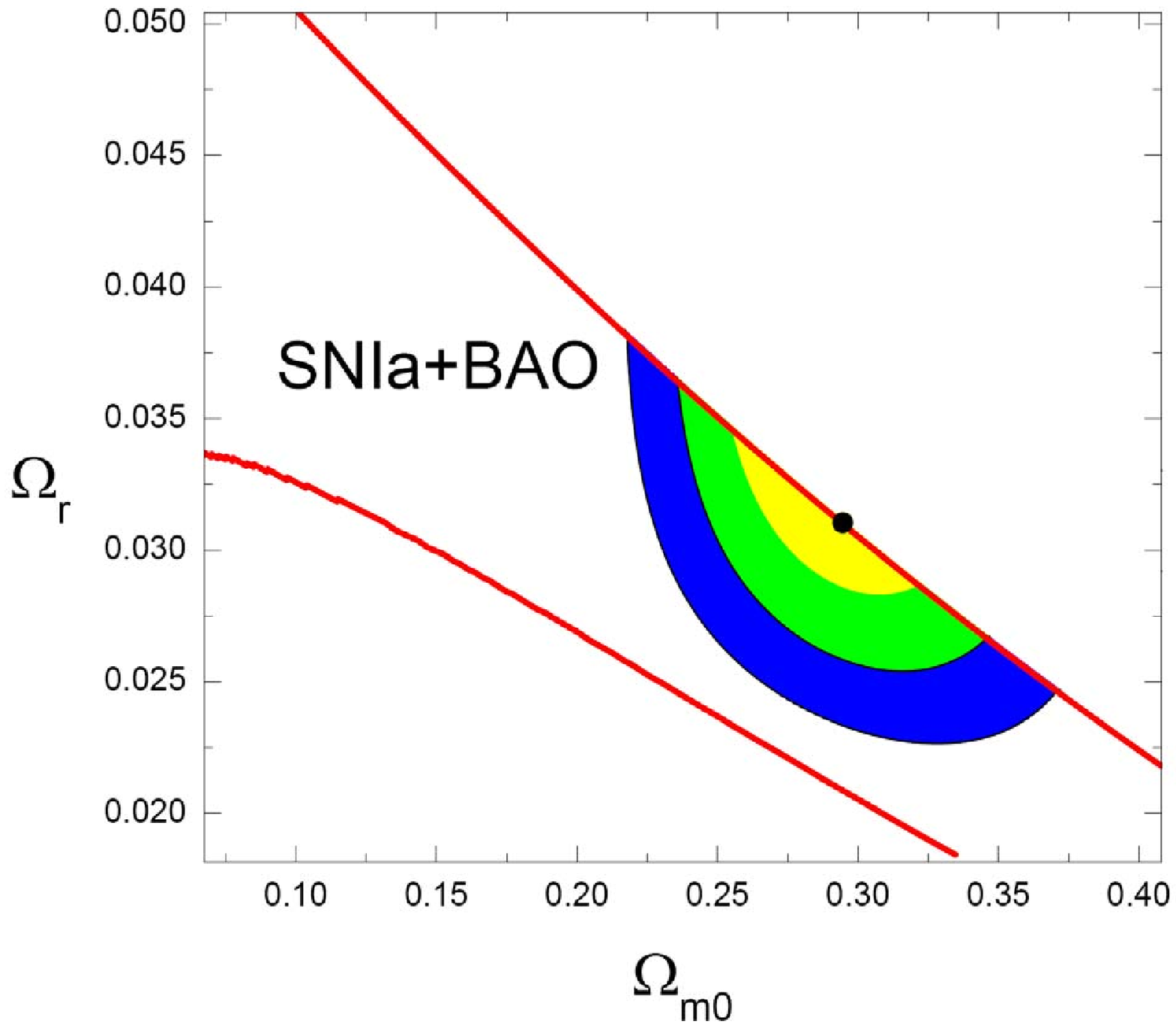}
\caption{Probability contours at 68.3\%, 95.4\% and 99.7\%
confidence levels for joint parameters: (a) ($\gamma $,
$\mu_{m0}$)-plane from SNIa+BAO data. The point shows the
best-fit. (b) For ($\Omega_{m0}$, $\Omega_{r}$)-plane. The upper
line is for the limit $\gamma=0$ while the lower line is due to
the limit set by Eq~(\ref{13}).\\\label{two}}
\end{figure}

\begin{table*}
\begin{center}
\caption{} \label{t2}
\begin{tabular}{|c|c|c|}
  \hline
  \hline
               & $\Omega_{m0}$ & $\chi^2_{min}$ \\
  \hline
  SNIa         & $0.24_{-0.03}^{+0.03}$ & 159.97 \\
  \hline
  SNIa+LSS     & $0.29_{-0.02}^{+0.02}$ & 162.92 \\
  \hline
  SNIa+LSS+H(z)&  $0.27_{-0.02}^{+0.02}$ & 174.04 \\
  \hline
  \hline
\end{tabular}
\end{center}
\end{table*}
An efficient way to reduce the degeneracies of the cosmological
parameters is to use the SNIa data in combination with the BAO
measurement from SDSS~\cite{10}. The acoustic signature in the
large scale clustering of galaxies yields additional test of
cosmology. Using a large sample of 46748 luminous red galaxies
covering 386 square degrees out to a redshif of $z=0.47$ from the
SDSS, Einstein et al~\cite{10} have found the model independent
BAO measurement which is described by the A parameter
\begin{eqnarray}
A & \equiv & \Omega_{m0}^{1/2}E(z_{\rm BAO})^{-1/3}\left[
\frac{1}{z_{\rm
BAO}}\int_0^{z_{\rm BAO}}\frac{dz'}{E(z')}\right]^{2/3} \nonumber \\
& = & \frac{\mu_0^{1/2}E(Z_{\rm BAO})^{-1/3}}{h_0}\left[
\frac{1}{z_{\rm BAO}}\int_0^{z_{\rm
BAO}}\frac{dz'}{E(z')}\right]^{2/3}~.
\end{eqnarray}
The measurement gives $A=0.469$$(n_s/0.98)^{-0.35}\pm0.017$ at
$z_{\rm BAO}=0.35$. The scalar spectral index is taken to be
$n_s=0.95$ through the three-year WMAP data. In our analysis, we
have investigated the joint statistics with the SN Ia data and the
BAO measuremen. The results are shown in Fig~\ref{two}~(a),~(b)
where we show the contours of 68.3\%, 95.4\% and 99.7\% confidence
level for $\mu_0$, $\gamma$ and $\Omega_{m0}$, $\Omega_{r}$
respectively. The fitted parameters with the $1\sigma$ errors are
shown in Table~\ref{t1}, where $h_0$ and $\alpha$ are obtained
from $\mu_0$ and $\gamma$.

\begin{figure}
\includegraphics[width=3.5in,height=2.9in]{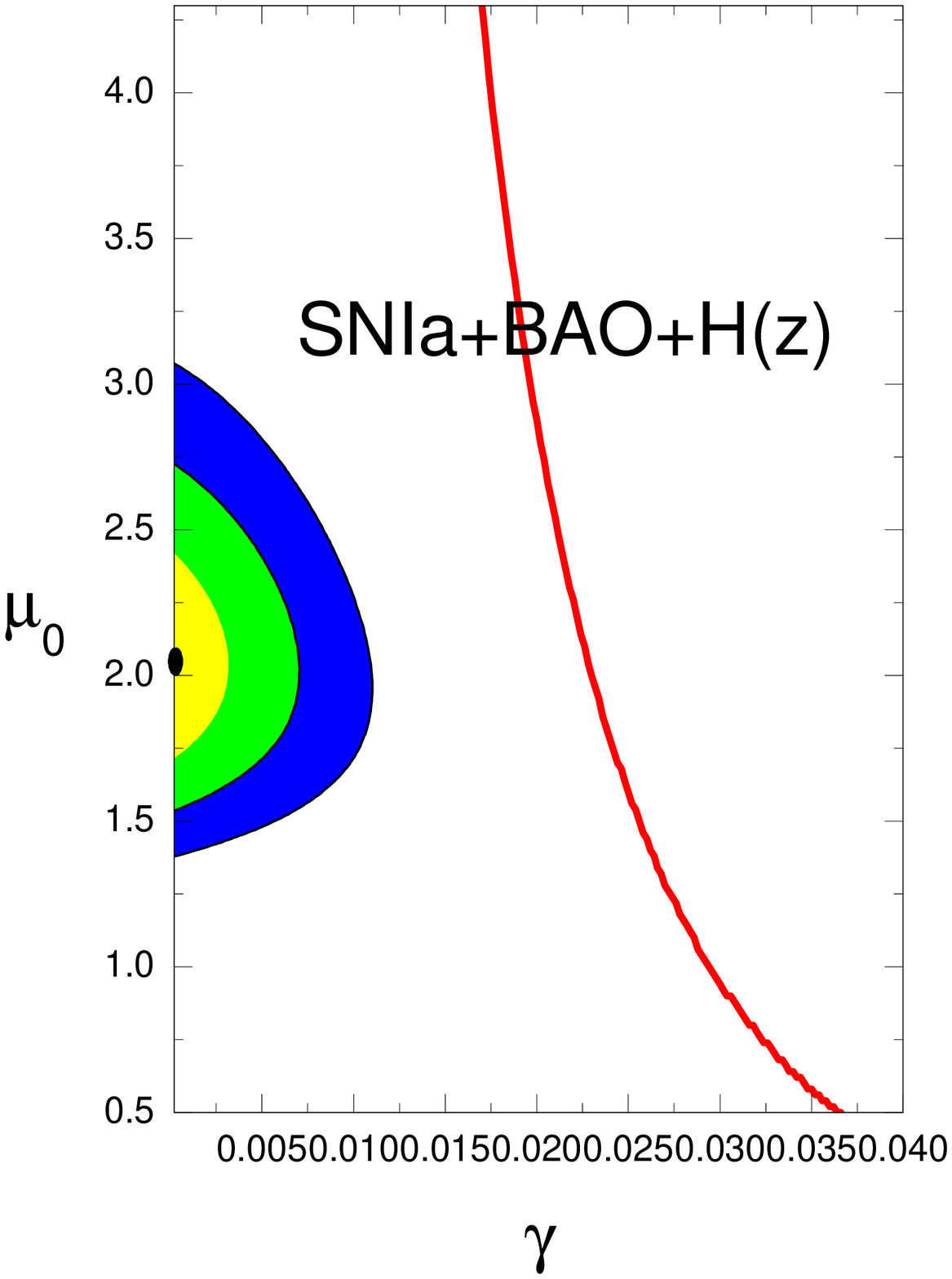}
\includegraphics[width=2.8in,height=2.8in]{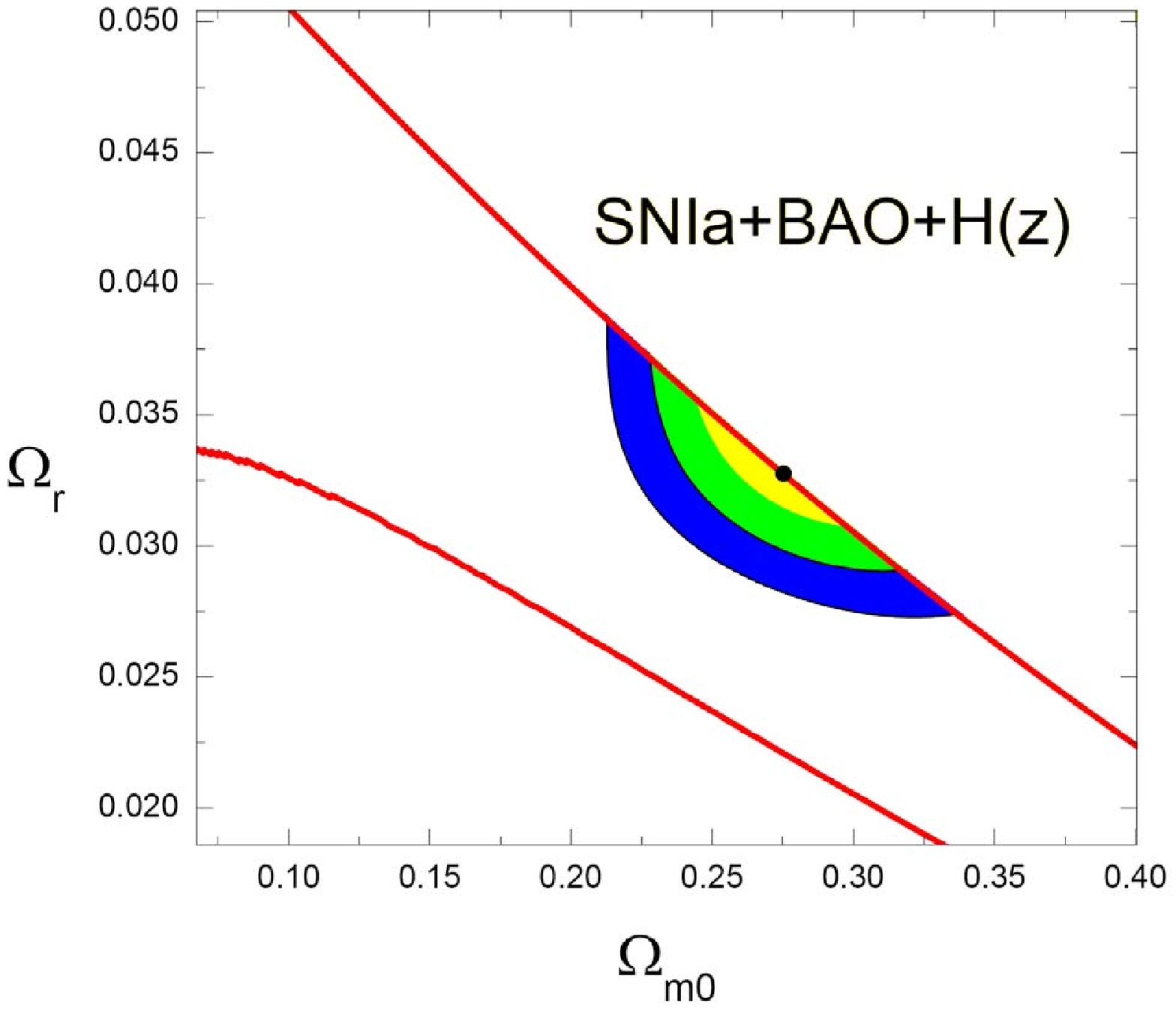}
\caption{Probability contours at 68.3\%, 95.4\% and 99.7\%
confidence levels for joint parameters: (a) ($\gamma $,
$\mu_{m0}$)-plane from SNIa+BAO+H(z) data. The point shows the
best-fit. (b) For ($\Omega_{m0}$, $\Omega_{r}$)-plane. The upper
line is for the limit $\gamma=0$ while the lower line is due to
the limit set by Eq~(\ref{13}).\\\label{three}}
\end{figure}

It is of interest to include the Hubble parameter data to
constrain our model. The Hubble parameter depends on the
differential age of the universe in terms of the redshift. In
contrast to standard candle luminosity distance, the Hubble
parameter is not integrated over. It persists fine structure which
is highly degenerated in the luminosity distance~\cite{14}.
Observed values of H(z)~\cite{15} can be used to place constraints
on the model of the expansion history of the universe by
minimizing the quantity
\begin{equation}
\chi^2_{H}(\gamma,\mu_0)=\sum_{i}\frac{\left|H_{obs}(z_i)-H_{th}(z_i)\right|^2}{\sigma_i^2}~.\label{18}
\end{equation}
The H(z) test on its own cannot provide tight constrain on the
model. It is interesting to combine the H(z) data with other
observational data to obtain tighter constraints on the
cosmological model. The result on the joint analysis H(z)+SNIa+BAO
is shown in Fig~\ref{three}~(a),~(b) respectively. $1\sigma$ range
parameters' spaces are listed in Table~\ref{t1}. It is interesting
to notice that errors of model parameters have been significantly
reduced.

For the sake of comparison, we have also done the fitting to
observations by using the pure DGP model where the GB correction
is absent. Results are shown in Table~\ref{t2}. Despite the big
difference in the single SNIa data fitting, in the combined
analysis SNIa+BAO,SNIa+BAO+H(Z), we can see that the GB correction
influence the universe evolution, however its effect is very
small. In Table~\ref{t3}, we include the fitting results for the
pure DGP model obtained in ~\cite{8},~\cite{11},~\cite{12}. For
comparison, we need to notice that $\Omega_{rc}=4\Omega_r$. Using
our best fit value of $\Omega_r$($\Omega_{rc}$),  we obtain
$r=H_0^{-1}\Omega_{rc}^{-1/2}=2.89H_0^{-1}$, which obeys the
inequality of (\ref{10}) due to the GB correction.
\begin{table*}
\begin{center}
\caption{} \label{t3}
\begin{tabular}{|c|c|c|c|c|c|}
  \hline
  \hline
  Model & Test & $\Omega_{m0}$ & $\Omega_{r_c}$ & $\Omega_K$ & ref \\
  \hline
  DGP& SNIa & $0.13_{-0.06}^{+0.06}$&$0.14_{-0.03}^{+0.03}$&$0.20_{-0.20}^{+0.20}$& \cite{8} \\
  \hline
  DGP& SNIa(new Gold)+CMB+SDSS+LSS & $0.28_{-0.03}^{+0.02}$&$0.13_{-0.01}^{+0.01}$&$-0.002_{-0.053}^{+0.064}$& \cite{8} \\
  \hline
  DGP& Gold+SNLS & $0.31_{-0.06}^{+0.07}$ & $0.23_{-0.03}^{+0.03}$ & & \cite{11} \\
  \hline
  DGP+GB& SNIa & $0.15_{-0.04}^{+0.11}$ & $0.12_{-0.01}^{+0.03}$ &  &  \\
  \hline
  DGP+GB & SNIa+LSS & $0.29_{-0.03}^{+0.03}$ & $0.13_{-0.01}^{+0.01}$ &  &  \\
  \hline
  DGP+GB & SNIa+LSS+H(z)& $0.27_{-0.02}^{+0.03}$& $0.13_{-0.01}^{+0.01}$& &\\
  \hline
  \hline
\end{tabular}
\end{center}
\end{table*}

If the value of the parameter $\gamma$ is not zero, we can find
the maximum redshift at which the universe started its existence
in our model. Taking the maximum value of $\gamma$ from
Table~\ref{t1} and using  relations (\ref{13}) and (\ref{14}) we
calculate   the $z_{max}$ allowed from observations in our model,
which we show in Table~\ref{t4}.

\begin{table*}
\begin{center}
\caption{} \label{t4}
\begin{tabular}{|c|c|c|c|}
  \hline
  \hline
    Test& $\gamma$ & $\mu_0$ &  $z_{max}$  \\
  \hline
   SNIa &$0.0278$ &$1.20_{-0.34}^{+0.70}$ &$1.11<z_{max}<1.76$   \\
   SNIa+LSS& $0.005$ & $2.26_{-0.42}^{+0.55}$  & $6.86<z_{max}<8.06$  \\
   SNIa+LSS+H(z)& $0.003$  & $2.03_{-0.33}^{+0.40}$  & $10.76<z_{max}<12.25$   \\
  \hline
  \hline
\end{tabular}
\end{center}
%\tablenotetext{$\alpha$ is in $H_0^{-2}$unit}
\end{table*}

Table IV is actually the result of fitting of our model to the
observational data (at late-time with small redshifts). The
central zero $\gamma$ corresponds to the infinite $z_{max}$, where
the GB effect can be neglected and then the model reduces to the
pure DGP. However, what is significant here is that the result is
quite sensitive to $\gamma$. Even small nonzero $\gamma$ will
cause dramatic change in the evolution of the universe. Table IV
takes $\gamma$ just at the edge of $1\sigma$ contour, which
illustrates that the effect of GB is significant and quite
possible. For the modified DGP model by including the GB
correction, it was argued that the combined induced gravity and GB
effects make the universe start at a finite maximum density and
finite pressure, but with infinite curvature\cite{7}. In other
words, the universe described in this model does not start at
$z=\infty$, but starts at a finite $z$ which is the maximum
redshift allowed in the model. $z_{max}$ in Table IV is this
maximum redshift when the combined induced gravity and GB effects
are taken into account. However, there is a question of the high
energy-early-time behaviour of the model. The maximum $z_{max}$
allowed from  Table IV is too small to accommodate the
conventional CMB formation at high redshifts even if we had the
technology to calculate such effects in our model (see \cite{cp}
for such an attempt). Therefore, to go to high redshifts region we
have to fine-tune $\gamma$ to very small values, making the
contribution of the GB term at late-times negligible.

In summary, in this work we have preformed a parameter estimation
of the  Induced Gravity model with a higher curvature GB term in
the bulk proposed in ~\cite{7}. We have analyzed data coming from
the most recent SN Ia sample, LSS observation and H(z)
measurement. The results show that the DGP model with the GB
correction is a viable candidate to explain the observed
acceleration of our universe. The value of the GB parameter
allowed by observation is very small giving only small effects to
the corrected DGP model. These correction effects are sensitive to
changes of the GB parameter. A nonzero value of $\gamma$ will
change significantly the cosmological evolution of our universe.
However, to make our model consistent with the conventional CMB
formation at high redshifts the GB parameter has to be fine-tuned
to very small values.

\section*{Acknowledgments}
This work was partially supported by NNSF of China, Ministry of
Education of China and Shanghai Educational Commission. E.P is
partially supported  by the European Union through the Marie Curie
Research and Training Network UniverseNet (MRTN-CT-2006-035863).
B.W. would like to acknowledge helpful discussions with Y. G.
Gong.


\begin{thebibliography}{99}

\bibitem{1} A. G. Riess et al, Astro. J. 116, 1009 (1998); S.
Perlmutter et al, Astrophy. J. 517, 565 (1999).

\bibitem{2}
%\cite{Kogan:1999wc}
%\bibitem{multigravity}
I.~I.~Kogan, S.~Mouslopoulos, A.~Papazoglou, G.~G.~Ross and
J.~Santiago,
%``A three three-brane universe: New phenomenology for the new  millennium?,''
Nucl.\ Phys.\ B {\bf 584}, 313 (2000) [arXiv:hep-ph/9912552];
%%CITATION = HEP-PH 9912552;%%
%\cite{Kogan:2000cv}
%\bibitem{Kogan:2000cv}
I.~I.~Kogan and G.~G.~Ross,
%``Brane universe and multigravity: Modification of gravity at large and small
%distances,''
Phys.\ Lett.\ B {\bf 485}, 255 (2000) [arXiv:hep-th/0003074];
%%CITATION = HEP-TH 0003074;%%
See, e.g.,
  R.~Gregory, V.~A.~Rubakov and S.~M.~Sibiryakov,
  %``Opening up extra dimensions at ultra-large scales,''
  Phys.\ Rev.\ Lett.\  {\bf 84}, 5928 (2000)
  [arXiv:hep-th/0002072];
  %%CITATION = HEP-TH 0002072;%%
  I.~I.~Kogan,
  %``Multi(scale)gravity: A telescope for the micro-world,''
  [arXiv:astro-ph/0108220];
  %%CITATION = ASTRO-PH 0108220;%%
  A.~Papazoglou,
  %``Brane-world multigravity,''
  [arXiv:hep-ph/0112159];
  %%CITATION = HEP-PH 0112159;%%
  A.~Padilla,
  %``Cosmic acceleration from asymmetric branes,''
  Class.\ Quant.\ Grav.\  {\bf 22}, 681 (2005)
  [arXiv:hep-th/0406157];
  %%CITATION = HEP-TH 0406157;%%
  A.~Padilla,
  %``Infra-red modification of gravity from asymmetric branes,''
  Class.\ Quant.\ Grav.\  {\bf 22}, 1087 (2005)
  [arXiv:hep-th/0410033].
  %%CITATION = HEP-TH 0410033;%%

\bibitem{3}
  G.~R.~Dvali, G.~Gabadadze and M.~Porrati,
  %``4D gravity on a brane in 5D Minkowski space,''
  Phys.\ Lett.\ B {\bf 485}, 208 (2000)
  [arXiv:hep-th/0005016].
  %%CITATION = HEP-TH 0005016;%%
  %\bibitem{DGP}
%\cite{Dvali:2000hr}
G.~R.~Dvali, G.~Gabadadze and M.~Porrati,
%``4D gravity on a brane in 5D Minkowski space,''
Phys.\ Lett.\ B {\bf 485}, 208 (2000) [arXiv:hep-th/0005016];
%%CITATION = HEP-TH 0005016;%%
G.~R.~Dvali and G.~Gabadadze,
%``Gravity on a brane in infinite-volume extra space,''
Phys.\ Rev.\ D {\bf 63}, 065007 (2001) [arXiv:hep-th/0008054];
%%CITATION = HEP-TH 0008054;%%
%\cite{Deffayet:2000uy}
C.~Deffayet,
%``Cosmology on a brane in Minkowski bulk,''
Phys.\ Lett.\ B {\bf 502}, 199 (2001) [arXiv:hep-th/0010186];
%%CITATION = HEP-TH 0010186;%%
%\cite{Deffayet:2001pu}
%\bibitem{Deffayet:2001pu}
C.~Deffayet, G.~R.~Dvali and G.~Gabadadze,
%``Accelerated universe from gravity leaking to extra dimensions,''
Phys.\ Rev.\ D {\bf 65}, 044023 (2002) [arXiv:astro-ph/0105068];
%%CITATION = ASTRO-PH 0105068;%%
%\cite{Lue:2005ya}
%\bibitem{Lue:2005ya}
A.~Lue,
%``The phenomenology of Dvali-Gabadadze-Porrati cosmologies,''
[arXiv:astro-ph/0510068].
%%CITATION = ASTRO-PH 0510068;%%


%\cite{Charmousis:2006pn}
\bibitem{DGPghosts}
C.~Charmousis, R.~Gregory, N.~Kaloper and A.~Padilla,
%``DGP specteroscopy,''
JHEP {\bf 0610}, 066 (2006) [arXiv:hep-th/0604086].
%%CITATION = HEP-TH 0604086;%%
%\cite{Padilla:2006eh}
%\bibitem{Padilla:2006eh}
A.~Padilla,
%``A short review of 'DGP Specteroscopy',''
[arXiv:hep-th/0610093].
%%CITATION = HEP-TH 0610093;%%
%\bibitem{koyama}
%\cite{Koyama:2005tx}
K.~Koyama,
%``Are there ghosts in the self-accelerating brane universe?,''
Phys.\ Rev.\ D {\bf 72}, 123511 (2005) [arXiv:hep-th/0503191];
D.~Gorbunov, K.~Koyama and S.~Sibiryakov,
%``More on ghosts in DGP model,''
Phys.\ Rev.\  D {\bf 73}, 044016 (2006) [arXiv:hep-th/0512097].
%%CITATION = HEP-TH 0512097;%%

\bibitem{strong}
%\cite{Rubakov:2003zb}
%\bibitem{Rubakov:2003zb}
V.~A.~Rubakov,
%``Strong coupling in brane-induced gravity in five dimensions,''
[arXiv:hep-th/0303125].
%%CITATION = HEP-TH/0303125;%%
%\cite{Luty:2003vm}
%\bibitem{Luty:2003vm}
M.~A.~Luty, M.~Porrati and R.~Rattazzi,
%``Strong interactions and stability in the DGP model,''
JHEP {\bf 0309}, 029 (2003) [arXiv:hep-th/0303116].
%%CITATION = JHEPA,0309,029;%%
%\cite{Dubovsky:2003pn}
%\bibitem{Dubovsky:2003pn}
S.~L.~Dubovsky and M.~V.~Libanov,
%``On brane-induced gravity in warped backgrounds,''
JHEP {\bf 0311}, 038 (2003) [arXiv:hep-th/0309131].
%%CITATION = JHEPA,0311,038;%%
%\cite{Smolyakov:2005hg}
%\bibitem{Smolyakov:2005hg}
M.~N.~Smolyakov,
%``The strong coupling effect and auxiliary fields in the DGP-model,''
Phys.\ Rev.\  D {\bf 72}, 084010 (2005) [arXiv:hep-th/0506020].
%%CITATION = PHRVA,D72,084010;%%



\bibitem{5}
See, e.g.,
  C.~Charmousis and J.~F.~Dufaux,
  %``General Gauss-Bonnet brane cosmology,''
  Class.\ Quant.\ Grav.\  {\bf 19}, 4671 (2002)
  [arXiv:hep-th/0202107];
  %%CITATION = HEP-TH 0202107;%%
  S.~Nojiri, S.~D.~Odintsov and S.~Ogushi,
  %``Friedmann-Robertson-Walker brane cosmological equations from the
  %five-dimensional bulk (A)dS black hole,''
  Int.\ J.\ Mod.\ Phys.\ A {\bf 17}, 4809 (2002)
  [arXiv:hep-th/0205187];
  %%CITATION = HEP-TH 0205187;%%
  S.~C.~Davis,
  %``Generalised Israel junction conditions for a Gauss-Bonnet brane world,''
  Phys.\ Rev.\ D {\bf 67}, 024030 (2003)
  [arXiv:hep-th/0208205];
  %%CITATION = HEP-TH 0208205;%%
  J.~E.~Lidsey and N.~J.~Nunes,
  %``Inflation in Gauss-Bonnet brane cosmology,''
  Phys.\ Rev.\ D {\bf 67}, 103510 (2003)
  [arXiv:astro-ph/0303168];
  %%CITATION = ASTRO-PH 0303168;%%
  K.~i.~Maeda and T.~Torii,
  %``Covariant gravitational equations on brane world with Gauss-Bonnet  term,''
  Phys.\ Rev.\ D {\bf 69}, 024002 (2004)
  [arXiv:hep-th/0309152];
  %%CITATION = HEP-TH 0309152;%%
  J.~F.~Dufaux, J.~E.~Lidsey, R.~Maartens and M.~Sami,
  %``Cosmological perturbations from brane inflation with a Gauss-Bonnet term,''
  Phys.\ Rev.\ D {\bf 70}, 083525 (2004)
  [arXiv:hep-th/0404161];
  %%CITATION = HEP-TH 0404161;%%
  T.~G.~Rizzo,
  %``Warped phenomenology of higher-derivative gravity,''
  JHEP {\bf 0501}, 028 (2005)
  [arXiv:hep-ph/0412087].
  %%CITATION = HEP-PH 0412087;%%

\bibitem{6}
  N.~E.~Mavromatos and E.~Papantonopoulos,
  %``Induced curvature in brane worlds by surface terms in string effective
  %actions with higher-curvature corrections,''
  [arXiv:hep-th/0503243].
  %%CITATION = HEP-TH 0503243;%%

\bibitem{7}
%\cite{Brown:2005ug}
%\bibitem{Brown:2005ug}
  R.~A.~Brown, R.~Maartens, E.~Papantonopoulos and V.~Zamarias,
  %``A late-accelerating universe with no dark energy - and no big bang,''
  JCAP {\bf 0511}, 008 (2005)
  [arXiv:gr-qc/0508116].
  %%CITATION = JCAPA,0511,008;%%


%\cite{Kofinas:2003rz}
\bibitem{Kofinas:2003rz}
  G.~Kofinas, R.~Maartens and E.~Papantonopoulos,
  %``Brane cosmology with curvature corrections,''
  JHEP {\bf 0310}, 066 (2003)
  [arXiv:hep-th/0307138].
  %%CITATION = JHEPA,0310,066;%%

\bibitem{bb} T. Koivisto, D. F. Mota,  Phys.Lett.B644, 104 (2007);
Phys.Rev.D75, 023518,(2007).


\bibitem{DGPobservations}
%\cite{Maartens:2006yt}
% \bibitem{Maartens:2006yt}
 R.~Maartens and E.~Majerotto,
  %``Observational constraints on self-accelerating cosmology,''
 Phys.\ Rev.\ D {\bf 74}, 023004 (2006) [arXiv:astro-ph/0603353];
M.C. Bento, O. Bertolami, M.J. Reboucas, N.M.C. Santos, Phys.Rev.
D73 (2006) 103521.

\bibitem{DGPobservations1}
 S.~Rydbeck, M.~Fairbairn and A.~Goobar,
  %``Testing the DGP model with ESSENCE,''
 JCAP {\bf 0705}, 003 (2007) [arXiv:astro-ph/0701495].
  %%CITATION = JCAPA,0705,003;%%

\bibitem{8}
%\cite{Movahed:2007ie}
% \bibitem{Movahed:2007ie}
 M.~S.~Movahed, M.~Farhang and S.~Rahvar,
  %``Recent observational constraints on the DGP modified gravity,''
   [arXiv:astro-ph/0701339].
    %%CITATION = ASTRO-PH/0701339;%%

\bibitem{11}
%\cite{Guo:2006ce}
% \bibitem{Guo:2006ce}
 Z.~K.~Guo, Z.~H.~Zhu, J.~S.~Alcaniz and Y.~Z.~Zhang,
  %``Constraints on the DGP Model from Recent Supernova Observations and Baryon %Acoustic Oscillations,''
 Astrophys.\ J.\ {\bf 646}, 1 (2006) [arXiv:astro-ph/0603632].
  %%CITATION = ASJOA,646,1;%%




\bibitem{12}
%\cite{Pires:2006rd}
%\bibitem{Pires:2006rd}
 N.~Pires, Z.~H.~Zhu and J.~S.~Alcaniz,
  %``Lookback time as a test for brane cosmology,''
   Phys.\ Rev.\ D {\bf 73}, 123530 (2006) [arXiv:astro-ph/0606689].
   %%CITATION = PHRVA,D73,123530;%%

\bibitem{b1}  A. Sheykhi, B. Wang and N. Riazi, Phys. Rev.  D {\bf 75}, 123513
(2007).

\bibitem{b2}
R.G.Cai, H.S.Zhang and A.Wang, Commun. Theor. Phys.  {\bf 44}, 948
(2005).

\bibitem{a1} K. Koyama and R. Maartens, JCAP {\bf 0601}, 016
(2006).

\bibitem{00}  M.C. Bento, O. Bertolami, M.J. Rebou?as, N.M.C.
Santos, Phys.Rev. D73 (2006) 103521.

\bibitem{9}
%\cite{Riess:2006fw}
% \bibitem{Riess:2006fw}
 A.~G.~Riess {\it et al.},
  %``New Hubble Space Telescope Discoveries of Type Ia Supernovae at $z > 1$:
   %Narrowing Constraints on the Early Behavior of Dark Energy,''
    [arXiv:astro-ph/0611572].
%%CITATION = ASTRO-PH/0611572;%%


\bibitem{10}
%\cite{Eisenstein:2005su}
% \bibitem{Eisenstein:2005su}
 D.~J.~Eisenstein {\it et al.} [SDSS Collaboration],
  %``Detection of the Baryon Acoustic Peak in the Large-Scale Correlation %Function of SDSS Luminous Red Galaxies,''
   Astrophys.\ J.\ {\bf 633}, 560 (2005) [arXiv:astro-ph/0501171].
 %%CITATION = ASJOA,633,560;%%






%\bibitem{13}
%\cite{Spergel:2006hy}
% \bibitem{Spergel:2006hy}
 %D.~N.~Spergel {\it et al.} [WMAP Collaboration],
  %``Wilkinson Microwave Anisotropy Probe (WMAP) three year results: %Implications for cosmology,''
 %[arXiv:astro-ph/0603449].
 %%CITATION = ASTRO-PH/0603449;%%


\bibitem{14}
%\cite{Wei:2006ut}
% \bibitem{Wei:2006ut}
 H.~Wei and S.~N.~Zhang,
  %``Observational $H(z)$ Data and Cosmological Models,''
   Phys.\ Lett.\ B {\bf 644}, 7 (2007) [arXiv:astro-ph/0609597];
 %%CITATION = PHLTA,B644,7;%%
%\cite{Wu:2006pe}
% \bibitem{Wu:2006pe}
 P.~Wu and H.~W.~Yu,
  %``Generalized Chaplygin gas model: Constraints
  % from Hubble parameter versus %redshift data,''
   Phys.\ Lett.\ B {\bf 644}, 16 (2007) [arXiv:gr-qc/0612055];
 %%CITATION = PHLTA,B644,16;%%
  %\cite{Wu:2007bv}
  % \bibitem{Wu:2007bv}
   P.~X.~Wu and H.~W.~Yu,
    %``Constraints on the unified dark energy-dark matter
    % model from latest %observational data,''
  JCAP {\bf 0703}, 015 (2007) [arXiv:astro-ph/0701446];
   %%CITATION = JCAPA,0703,015;%%
%\cite{Xu:2007nj}
% \bibitem{Xu:2007nj}
 L.~I.~Xu, C.~W.~Zhang, B.~R.~Chang and H.~Y.~Liu,
  %``Reconstruction of Deceleration Parameters from Recent Cosmic %Observations,''
 [arXiv:astro-ph/0701519];
  %%CITATION = ASTRO-PH/0701519;%%
%\cite{Kurek:2007tb}
%\bibitem{Kurek:2007tb}
 A.~Kurek and M.~Szydlowski,
  %``The LambdaCDM model on the lead -- a Bayesian cosmological models %comparison,''
  [arXiv:astro-ph/0702484];
  %%CITATION = ASTRO-PH/0702484;%%
%\cite{Zhang:2007uh}
% \bibitem{Zhang:2007uh}
 H.~Zhang and Z.~H.~Zhu,
  %``Natural Phantom Dark Energy, Wiggling Hubble Parameter $H(z)$ and Direct %$H(z)$ Data,''
  [arXiv:astro-ph/0703245].
 %%CITATION = ASTRO-PH/0703245;%%



\bibitem{15}
%\cite{Simon:2004tf}
% \bibitem{Simon:2004tf}
 J.~Simon, L.~Verde and R.~Jimenez,
  %``Constraints on the redshift dependence of the dark energy potential,''
  Phys.\ Rev.\ D {\bf 71}, 123001 (2005) [arXiv:astro-ph/0412269].
   %%CITATION = PHRVA,D71,123001;%%

\bibitem{cp}  S. Tsujikawa, M. Sami and R. Maartens,  Phys. Rev.  D {\bf 70}, 063525
(2004).

\end{thebibliography}
\end{document}